\def\spose#1{\hbox to 0pt{#1\hss}}
\def\approxlt{\mathrel{\spose{\lower 3pt\hbox{$\sim$}}
	\raise 2.0pt\hbox{$$<$$}}}
\def\approxgt{\mathrel{\spose{\lower 3pt\hbox{$\sim$}}
	\raise 2.0pt\hbox{$>$}}}

\def\multleft#1{\hbox to size{\vbox {\halign {\lft{##}\cr #1}}\hfill}\par}
\def\multright#1{\hbox to size{\vbox {\halign {\rt{##}\cr #1}}\hfill}\par}

\def\today{\ifcase\month\or January\or February\or March\or April\or May\or
      June\or July\or August\or September\or October\or November\or December\fi
      \space\number\day, \number\year}
\def\$<${\thinspace}
\def\s{\hbox{\phantom{5}}}	

\def\boxit#1{\vbox{\hrule\hbox{\vrule\kern3pt\vbox{\kern3pt
          #1 \kern3pt}\kern3pt\vrule}\hrule}}

\def\cm{{\rm\thinspace cm}}

\def\erg{{\rm\thinspace erg}}

\def\keV{{\rm\thinspace keV}}
\def\km{{\rm\thinspace km}}

\def\Mpc{{\rm\thinspace Mpc}}

\def\s{{\rm\thinspace s}}

\def\ergpcmsqps{\hbox{$\erg\cm^{-2}\s^{-1}\,$}}

\def\ergps{\hbox{$\erg\s^{-1}\,$}}

\def\kmps{\hbox{$\km\s^{-1}\,$}}

\def\pcmsq{\hbox{$\cm^{-2}\,$}}

\def\psqcm{\hbox{$\cm^{-2}\,$}}

\def\kmpspMpc{\hbox{$\kmps\Mpc^{-1}$}}

\def\ergpspcubMpc{erg~s$^{-1}$~Mpc$^{-3}$}

\documentstyle[psfig]{mn}
\begin{document}
\hsize=6truein

\title{Fitting the spectrum of the X-ray background: the effects of high metallicity absorption}

\author[]
{\parbox[]{6.in} {R.J.~Wilman and A.C.~Fabian \\ 
\footnotesize
Institute of Astronomy, Madingley Road, Cambridge CB3 0HA \\ }}

\maketitle

\begin{abstract}
Recent work by Risaliti et al.~(1999) suggests that more than half of all 
Seyfert 2 galaxies in the local universe are Compton-thick ($N_{\rm{H}}>10^{24}$\psqcm). This has implications for AGN synthesis models for the X-ray background (XRB), the flexibility of which for the inclusion of large numbers of high-$z$ type 2 sources we examine here. We highlight the importance of Compton down-scattering in determining the individual source spectra and the fit to the XRB spectrum, and demonstrate how parameter space `opens up' considerably if a super-solar iron abundance is assumed for the absorbing material. This is illustrated with a model which satisfies the present constraints, but which predicts substantial numbers of type 2 sources at the faint flux levels soon to be probed for the first time by the Chandra and XMM missions. We demonstrate also how a strong negative K-correction facilitates the detection of sources with $10^{\sim24} \leq N_{\rm{H}} \leq 10^{25}$\psqcm out to the highest redshifts at which they could plausibly exist. 
\end{abstract}


\section{INTRODUCTION}
There have been numerous attempts to account for the origin of the cosmic X-ray background (XRB) with populations of active galactic nuclei (AGN) integrated over cosmic time (Setti \& Woltjer 1989; Madau et al.~1994; Celotti et al.~1995; Comastri et al.~1995). In this case, Fabian \& Iwasawa (1999) have recently shown that the XRB sources plausibly account for all of the accretion power in the Universe, and that most of it is intrinsically absorbed. The XRB is thus an important probe of the evolution of the black hole population with cosmic epoch. It is of interest to know how this relates to the evolution of star formation in galaxies, which work at sub-mm wavelengths is beginning to constrain (eg. Hughes et al.~1998).  

Within the framework of unified models for AGN, XRB synthesis models require as inputs the X-ray spectra of the unabsorbed (type 1) and absorbed (type 2) sources, and the local X-ray luminosity function (XLF) for a particular class of source together with its cosmological evolution. It has been common practice to adopt an XLF for the type 1 sources, and to convolve it with an $N_{\rm{H}}$ distribution prescribing the relative numbers of sources with different amounts of absorption. The form of the $N_{\rm{H}}$ distribution is, however, the major uncertainty in current models; the earliest works assumed it to be independent of both luminosity and redshift, and consistent with the poorly-constrained local estimates of the Seyfert 2/Seyfert 1 number ratio. The latter lies between $\sim$1.6--3 according to Huchra \& Burg~(1992) for an optically-selected Seyfert sample, but other studies constrain it to lie between 2 and 10 (Osterbrock \& Shaw 1988; Goodrich, Veilleux \& Hill 1994). More recently, there have been attempts to determine the $N_{\rm{H}}$ distribution directly: from a large sample of Seyfert 2 galaxies, Bassani et al.~(1998) found a mean $\log N_{\rm{H}}$ of 23.5 with 23--30 per cent of sources having $\log N_{\rm{H}}>24$. Their distribution is, however, biased in favour of low $N_{\rm{H}}$ values, as they include all Seyfert 2s for which a `reliable' hard X-ray spectrum is available in the literature. Risaliti et al.~(1999) overcame this and other selection biases by using the [OIII]$\lambda5007$ flux to select a sub-sample of Seyfert 2 galaxies from the sample of Maiolino \& Rieke~(1995); they found an $N_{\rm{H}}$ distribution even more strongly dominated by heavily absorbed sources, with 75 per cent having $\log N_{\rm{H}}>23$ and about half having $\log N_{\rm{H}}>24$. They found no correlation between $N_{\rm{H}}$ and the intrinsic luminosity of the nuclear source.

Attempts to incorporate some of this new information into the synthesis models have recently been made (Gilli et al.~1999a). We emphasize that the above $N_{\rm{H}}$ distributions are determined locally, and that the situation at higher redshifts may be quite different. To this end, we examine in the present work how much flexibility there is within the models for accommodating type 2 sources in substantially larger numbers at high redshift, at flux levels within reach of the forthcoming generation of X-ray observatories (Chandra and XMM). In so doing, we draw attention to the role played by Compton down-scattering in determining the spectra of the individual sources and hence the quality of the fit to the XRB spectrum. Any model which includes large numbers of Compton-thick sources must clearly incorporate this feature; it was omitted by Comastri et al.~(1995), and (as pointed out by Celotti et al.~1995) handled erroneously by Madau et al.~(1994). Only Celotti et al.~(1995) appear to have treated it properly, and they found a very restricted range of parameter space for acceptable models. We demonstrate that the quality of the fit to the XRB spectrum can be improved, and the available parameter space enlarged, by positing a super-solar metallicity for the absorbing material.

We adopt the cosmological parameters $H_{\rm{0}}=50$\kmpspMpc~and $q_{\rm{0}}=0.5$ throughout.

\section{THE MODEL}

\subsection{The AGN XLF and evolution}
In common with Madau et al.~(1994) and Celotti et al.~(1995) we adopt the Piccinotti et al.~(1982) 2--10\keV~XLF for local type 1 sources: $n(L_{44})=n_{0}L_{44}^{-2.75}$, where $L_{44}$ is the 2--10\keV~luminosity in units of 10$^{44}$\ergps and $n_{0}=2.68 \times 10^{-7}$Mpc$^{-3} (10^{44}\ergps)^{-1}$, between $L_{X,min}=7 \times 10^{42}$\ergps~and $L_{X,max}=100L_{X,min}$. Others (eg. Gilli et al.~1999a, 1999b, Comastri et al.~1995), use the 0.3--3.5\keV~XLF for broad line objects (QSOs and Seyfert 1s) derived from ROSAT and (Einstein) EMSS data (Jones et al.~1997 and references therein). Whilst the ROSAT+EMSS XLF extends to higher luminosities than that of Piccinotti et al.~(1982), the integrated number density derived from the latter is larger by an order of magnitude, showing that there are many 2--10\keV~sources absent from the 0.3--3.5\keV~band. It is unlikely that this discrepancy can be wholly ascribed to the contamination of the Piccinotti et al. sample by type 2 sources (of which there are 7, compared with 18 of type 1).

We adopt pure luminosity evolution of the form $(1+z)^{p}$ out to a cut-off redshift $z_{cut}$, with no further evolution thereafter out to a redshift $z_{max}$. This is consistent with the findings of Jones et al.~(1997) who derive $z_{cut}\sim1.4$ and $p\sim3$. We note that Miyaji et al.~(1998) have recently determined the XLF in the {\em observed} 0.5--2.0\keV~band for all AGN (ie. not just the type 1 objects) by combining the existing ROSAT survey data. They propose a luminosity-dependent density evolution model, but do not provide a break-down according to source classification, so the effects of intrinsic absorption render their XLF of limited use for synthesis models.

\subsection{Source spectra}
For the primary type 1 X-ray spectrum, we follow Madau et al.~(1994) and assume the form $I(E) \propto E^{-0.9}exp(-E/E_{c})$ with $E_{c}$ fixed at 360\keV. To this is added a component reflected from an accretion disc subtending $2\pi$~sr at the source, for a fixed inclination angle of $60^{\circ}$. Type 2 spectra are generated by feeding this primary spectrum to a Monte-Carlo code which computes the transmission through a homogeneous sphere of cold material with a radius given by the column density $N_{\rm{H}}$. Both photoelectric absorption from neutral material (of solar composition unless explicitly stated otherwise) and Compton down-scattering are modelled, the latter being described by the Klein-Nishina cross-section. Investigations with the obscuring material in the form of shells and slabs showed that the output spectra are largely independent of the assumed geometry. Finally, a component equal to 2 per cent of the primary incident power is added to the type 2 spectra to represent the flux scattered into the line of sight by electrons in the warm, ionized medium above the axis of the obscuring torus. The Monte-Carlo spectra closely match those generated independently by Celotti et al.~(1995) (A.~Celotti, private communication).  

We note that Celotti et al.~(1995) assumed the sources to be compact and in thermal pair equilibrium, as a result of which the cut-off energy $E_{c}$ evolves with source luminosity. For simplicity, we do not model this here as it can be shown that if a range of $E_{c}$ values is employed, the mean must be $\sim300$\keV~in order to match the slope of the observed XRB (Gruber et al.~1992) between 30 and 100\keV~(which is almost entirely intrinsically unabsorbed). 

The $N_{\rm{H}}$ distribution is parameterised as $N(\log N_{\rm{H}}) \propto (\log N_{\rm{H}})^{\beta}$ from $N_{\rm{H},min}=10^{20}$\psqcm~to $N_{H,\star}=10^{22}$\psqcm~and from $N_{\rm{H},\star}$ to $N_{\rm{H},max}$ for sources of types 1 and 2, respectively. In a departure from previous works, we investigate the effects of a redshift-dependent $\beta$, but do not incorporate any explicit luminosity dependence. The results of Risaliti et al.~(1999) are consistent with the latter assumption (at least locally), and with $\beta \sim 8$. 

\begin{figure}
\psfig{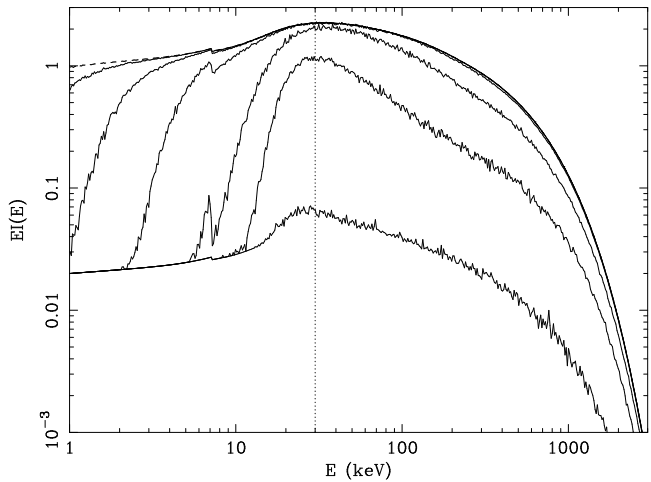}
\caption{\normalsize For the primary input spectrum described in the text (dashed line), the solid lines are the type 2 spectra produced by the Monte-Carlo code (plus the 2 per cent component scattered by warm ionized material) for the column densities (from left to right) $N_{\rm{H}}=10^{21.25}, 10^{22.25}, 10^{23.25}, 10^{24.25}, 10^{24.75}$ and $10^{25.25}$\psqcm. Note that all the spectra peak at $E\leq30$\keV, as the position of dotted vertical line makes clear.}
\label{fig:mcspectraFe1}
\end{figure}

\begin{figure}
\psfig{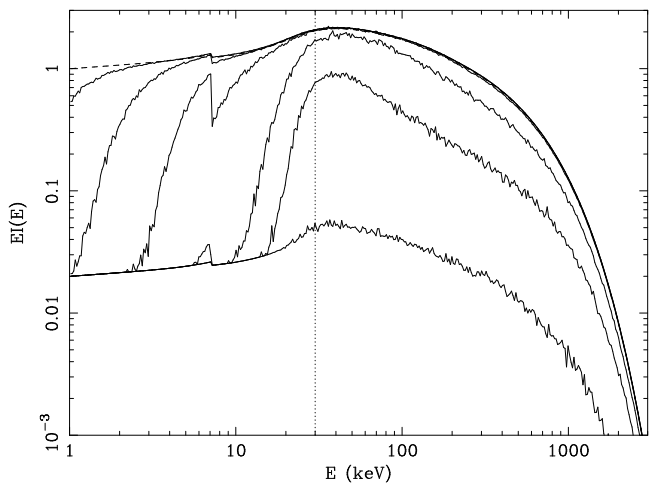}
\caption{\normalsize As in Fig.~\ref{fig:mcspectraFe1}, but for an iron abundance of $5\times$~solar. Note how the peaks in the spectra have been shifted to $E>30$\keV, which is crucial for reproducing the observed peak in the XRB spectrum at 30\keV.}
\label{fig:mcspectraFe5}
\end{figure}

\begin{figure}
\psfig{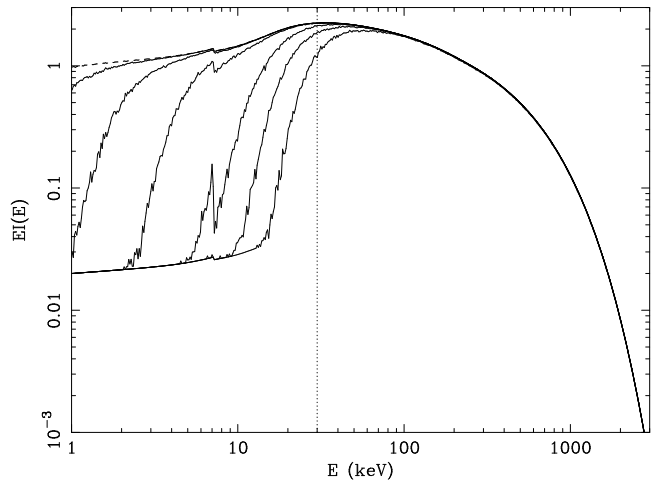}
\caption{\normalsize As in Fig.~\ref{fig:mcspectraFe1} but with Compton down-scattering `turned off'; note how the spectral peaks of the Compton-thick sources are shifted well above 30\keV, thus creating extra parameter space for XRB synthesis models which neglect this important effect.}
\label{fig:photoelspectra}
\end{figure}

In Fig.~\ref{fig:mcspectraFe1} we show the spectra generated by our Monte-Carlo code for various $N_{\rm{H}}$ values, assuming solar metallicity for the absorbing material. It is clear that all the spectra peak in $EI(E)$ at $E\leq30$\keV. Thus, regardless of the details of the $N_{\rm{H}}$ distribution and of the form of the cosmological evolution, there is little scope for generating an XRB spectrum peak at the observed energy of 30\keV, especially if large numbers of high redshift sources are included. Within the formalism of the present synthesis model, the only means of `opening up' the parameter space is to posit a super-solar metallicity for the absorbing material. Specifically, by raising the iron abundance the photoelectric absorption can be increased whilst not substantially affecting the amount of Compton down-scattering, thereby shifting the peaks in the $EI(E)$ spectra to higher energies (the positions of the peaks are relatively insensitive to the value of $E_{c}$). This is illustrated in Fig.~\ref{fig:mcspectraFe5} for an iron abundance of $5\times$solar. If Compton down-scattering is `turned off' so that photoelectric absorption remains the only source of opacity, this particular difficulty does not arise, as Fig.~\ref{fig:photoelspectra} shows.

\section{RESULTS}
We discuss here the model fits to both the XRB spectrum and the source counts ($\log N-\log S$) in the several bands, for various values of the evolutionary parameters $z_{cut}$, $z_{max}$ and $p$, and different $N_{\rm{H}}$ distributions. We parameterise the latter with the following redshift-dependence for $\beta$:\\
$\beta=\beta_{1}$ for $0 \leq z < 0.5$ \\
$\beta=\beta_{2}$ for $0.5 \leq z < 1.0$ \\
$\beta=\beta_{3}$ for $1.0 \leq z < 2.0$ \\
$\beta=\beta_{4}$ for $2.0 \leq z < 3.0$ \\
$\beta=\beta_{5}$ for $3.0 \leq z < z_{max}$ \\ \\
We take $N_{\rm{H},max}=10^{25}$\psqcm, as Figs.~\ref{fig:mcspectraFe1} and~\ref{fig:mcspectraFe5} show that there is essentially no direct flux (and hence little contribution to the XRB) from more heavily obscured sources (but Fig.~\ref{fig:photoelspectra} shows that this is not so when Compton down-scattering is neglected). Acceptable fits to the observed XRB spectrum as parameterised above 3\keV~by Gruber et al.~(1992) were defined as satisfying the criteria given by Celotti et al.~(1995). As stated therein, the model spectrum is first re-normalized to the observed XRB at 30\keV~by between $\sim 0.7$ and 1.5, to take into account uncertainty in the normalization of the type 1 XLF, as set by the assumed value of $N_{\rm{H},\star}$ and the level type 2 contamination in the Piccinotti et al.~(1982) XLF. 

\subsection{Models with solar iron abundance}
With solar iron abundance and $\beta_{1-5}=8$, acceptable models can be found for z$_{cut} \simeq$1.2--1.4, $p \simeq$2.5--2.9 and z$_{max} \simeq$3.5--4.0 (ie. within the 68 per cent confidence limits derived by Jones et al.~1997). There is no freedom for the inclusion of larger numbers of type 2 sources at high redshift. The predicted local emissivities in the 2--10\keV~band lie at or just beyond the upper end of the range $\sim(3.5-4.8) \times 10^{38}$h$_{50}~$\ergpspcubMpc~derived by Miyaji et al.~(1994), from the cross-correlation of the XRB measured with HEAO-1 and IRAS galaxies. The models resolve $\simeq 95-100$ per cent of the 2--10 \keV~XRB intensity measured by Marshall et al.~(1980), which can be compared with an estimated contribution from clusters of galaxies of $\simeq 4.7$ per cent. We obtained the latter figure by integrating the Ebeling et al.~(1997) cluster XLF over $10^{42}\leq L(2-10\keV)\leq 10^{46}$\ergps, assuming the David et al.~(1993) $L_{\rm{X}}-T$ relation for a thermal bremsstrahlung spectrum and no cosmological evolution out to a maximum redshift z=4. The derived $\log N-\log S$ relation in the 2--10\keV~band is consistent with the constraints from Piccinotti et al.~(1982), Butcher et al.~(1997) and with the recent measurements down to $S(2-10\keV)=10^{-13}$\ergpcmsqps~from the ASCA Large Sky Survey (Ueda et al.~1999).

\subsection{Models with super-solar iron abundance}
\begin{figure}
\psfig{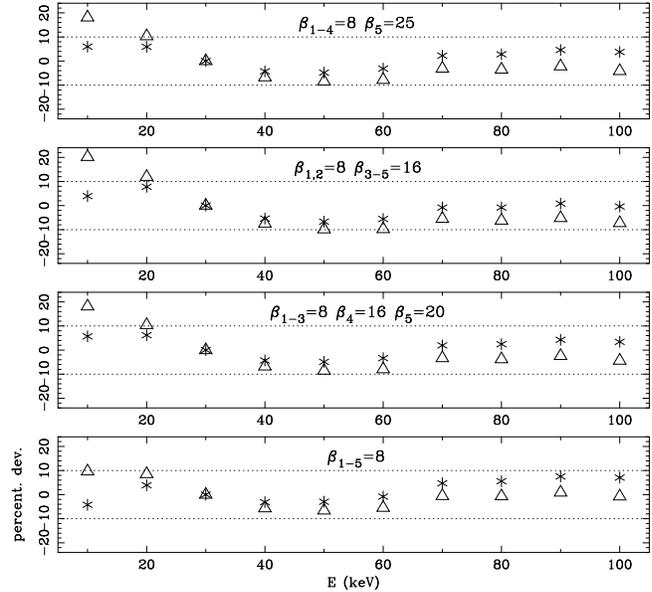}
\caption{\normalsize For iron abundances of solar (triangles) and 5 times solar (asterisks), each of the panels shows, for a different $N_{\rm{H}}$ distribution, the percentage deviation of the model from the observed (Gruber 1992) XRB spectrum at energies between 10 and 100\keV, after normalization at 30\keV. In each panel, the models shown provide the best-fit to the spectrum for $z_{cut}\simeq1.3$, $z_{max}\simeq4.0$ and $p\simeq2.5$. The $\pm10$~per cent limits for acceptable models and the $\beta$ parameters of the $N_{\rm{H}}$ distributions are as indicated. There is a larger range of permitted $N_{\rm{H}}$ distributions with the higher iron abundance.}
\label{fig:deviation}
\end{figure}

As discussed in section~2, much more parameter space is available to satisfactory models if the metallicity of the absorbing material is increased. In particular, there is scope for including substantially more type 2 sources at high-redshift. This is illustrated in Fig.~\ref{fig:deviation} where we show the percentage deviation of the model from the Gruber~(1992) fit to the observed spectrum for various $N_{\rm{H}}$ distributions and for two different iron abundances. We consider models constructed using an iron abundance of 5 times solar, which is perhaps an extreme value but which should bracket the likely range. Rather than provide an extensive tour of parameter space, we illustrate what is possible with an arbitrarily chosen model for which $\beta_{1-3}=8$, $\beta_{4,5}=16$, $z_{cut}=1.3$, $z_{max}=4.0$ and $p=2.5$. After renormalization at 30\keV~(by a factor of 0.93), this model resolves 90 per cent of the Marshall et al.~(1980) 2--10\keV~XRB intensity, with a 2--10\keV~emissivity of $4.30 \times 10^{38}$h$_{50}~$\ergpspcubMpc. [Formally acceptable fits with some of the extreme $\beta$ values of Fig.~\ref{fig:deviation} can be obtained with a solar iron abundance by substantially increasing the value of $E_{\rm{c}}$ and renormalizing appropriately; such models, however, yield XRB spectra with slopes which are too flat above 30\keV, as mentioned in section~2.2].

\begin{figure}
\psfig{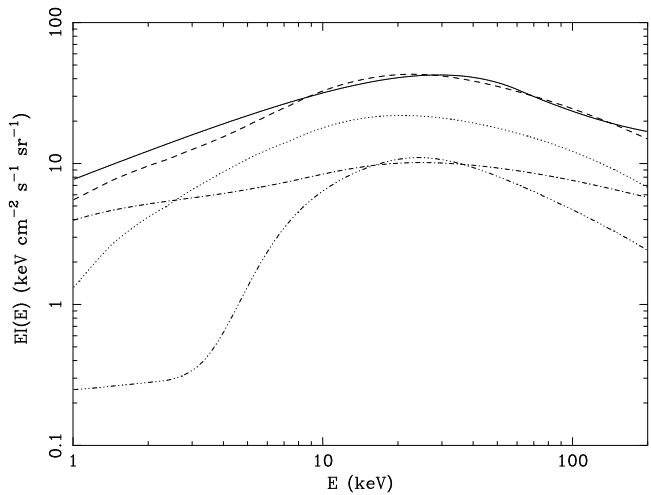}
\caption{\normalsize Comparison of the XRB spectrum of the model described in the text for an iron abundance of $5\times$~solar (dashed line) with the analytical fit of Gruber et al.~(1992) to the observed spectrum (solid line). The contributions to the model from type 1 sources (dot-dashed line), and the Compton-thin (dotted line) and Compton-thick (three dots-dashed line) type 2 sources are as shown.}
\label{fig:spmfit}
\end{figure}

\begin{figure}
\psfig{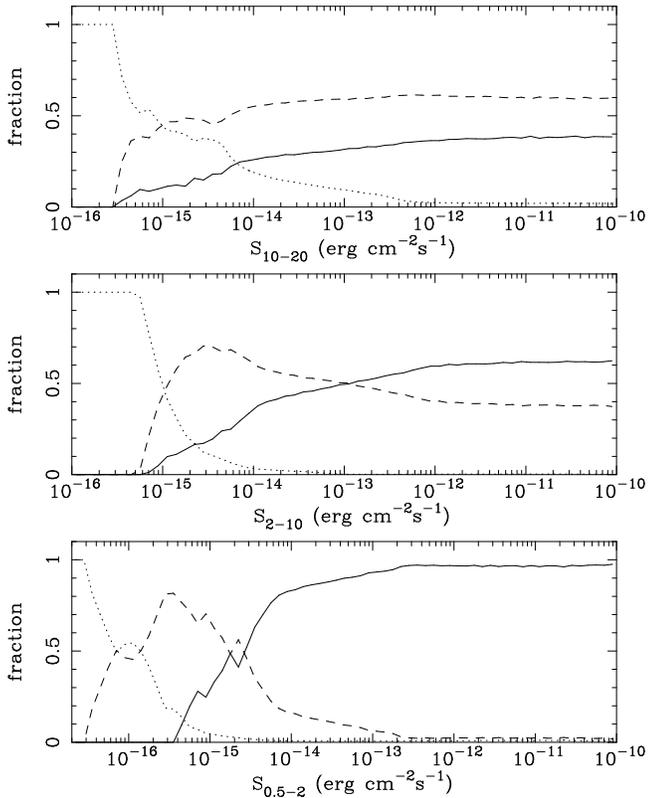}
\caption{\normalsize The fraction of type 1 (solid line), Compton-thin (dashed line) and Compton-thick (dot-dashed line) type 2 sources, are shown as functions of $S(10-20\keV)$  $S(2-10\keV)$ and $S(0.5-2.0\keV)$ for the model described in section~3.2, with an Fe abundance of $5\times$~solar.}
\label{fig:fraction}
\end{figure}

Fig.~\ref{fig:spmfit} shows the fit to the Gruber et al.~(1992) parameterization of XRB spectrum, along with the contributions of the different classes of source. Figs.~\ref{fig:lognlogshard} and~\ref{fig:lognlogssoft} show the model predicted source counts in the hard (2--10\keV) and soft (0.5--2\keV) bands respectively, along with the current observational constraints. Fig.~\ref{fig:lognlogshard} saturates at a flux level almost two orders of magnitude lower than the models proposed by Celotti et al.~(1995), in part because the latter authors used a lower $z_{max}$=2.6 (and a correspondingly higher $z_{cut}$=2.0). Compton-thin type 2 sources are dominant in the interval $10^{-15} \leq S(2-10\keV) \leq 10^{-14}$\ergpcmsqps and the surface density of Compton-thick sources in $10^{-16}\leq S(2-10\keV) \leq 10^{-15}$\ergpcmsqps~is $\sim1600$~deg$^{-2}$. This is of interest as the forthcoming Chandra and XMM missions will probe down to 2--10\keV~flux levels of $\sim10^{-16}$ and $\sim10^{-15}$\ergpcmsqps, respectively. For comparison with the preliminary BeppoSAX High Energy Large Area Survey (HELLAS) results (Comastri et al.~1999), we show in Fig.~\ref{fig:lognlogs5to10} the source counts in the 5--10\keV~band; the discrepancy at intermediate fluxes may be due to the contribution of non-AGN sources or to residual calibration problems, as described by the latter authors. Because of its potential relevance to future missions, we show in Fig.~\ref{fig:lognlogs10to20} the source counts for the 10--20\keV~band. Fig.~\ref{fig:fraction} shows the fractions of sources of each type as a functions of $S(10-20\keV)$, $S(2-10\keV)$ and $S(0.5-2.0\keV)$. 

The discrepancy between the model-predicted and observed source counts in the 0.5--2.0\keV~band perhaps indicates that our modelling of the soft X-ray spectra of the sources is unrealistic. Indeed, it is partly because there are large uncertainties in the spectral shape of the soft X-ray emission in AGN that we follow Celotti et al.~(1995) in neglecting the soft X-ray band when assessing the quality of the fit to the XRB spectrum. There is some scope for adjusting the model counts at faint 0.5--2.0\keV~fluxes in Fig.~\ref{fig:lognlogssoft} by changing the scattering fraction for the type 2 sources from the figure of 2 per cent used here. That there might be some dispersion in this quantity is suggested by ROSAT HRI observations of some hyperluminous IRAS galaxies, where the upper limits on the 0.1--2.4\keV~fluxes imply scattering fractions below $\sim 0.5$~per cent (Fabian et al.~1996; Wilman et al.~1998).

\subsection{Detecting Compton-thick sources at high redshift}
\begin{figure}
\psfig{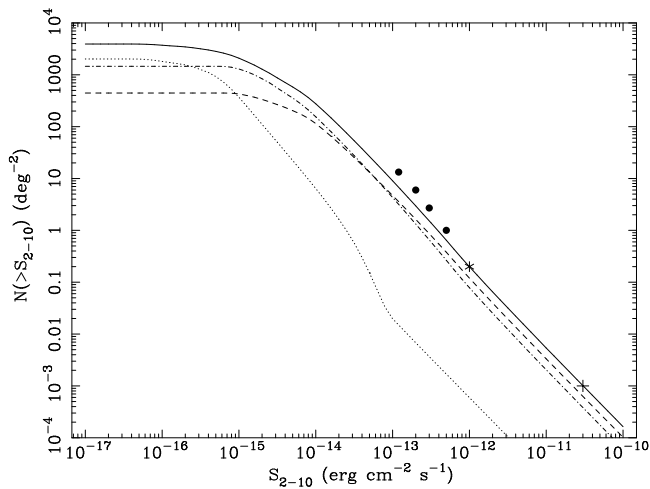}
\caption{\normalsize Predicted $\log N - \log S$ in the 2--10\keV~band of the model described in section~3.2. The solid line shows the sum of all the source counts, the dashed line the type 1 sources with $20<\log N_{\rm{H}}<22$, the dash-dot line type 2 sources with $22<\log N_{\rm{H}} <24$ and the dotted line those sources with $24<\log N_{\rm{H}} <25$. The cross is the HEAO-1 A2 AGN surface density from Piccinotti et al.~(1982), the asterisk the result of GINGA fluctuation analysis after the removal of the cluster contribution (Butcher et al.~1997) and the filled circles represent the results of the ASCA Large Sky Survey (Ueda et al.~1999).}
\label{fig:lognlogshard}
\end{figure}

\begin{figure}
\psfig{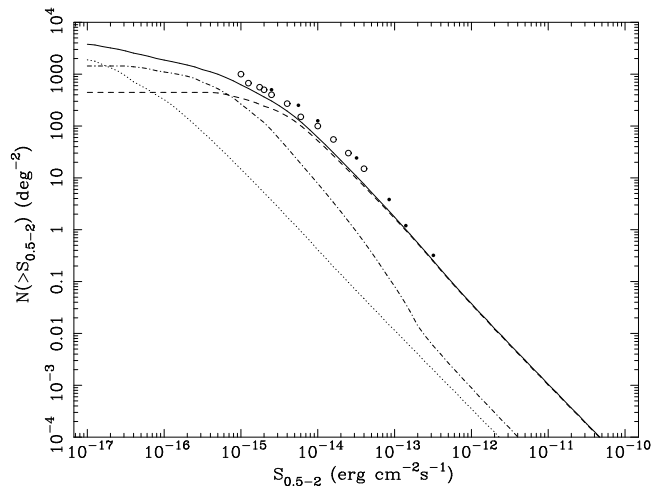}
\caption{\normalsize As in Fig.~\ref{fig:lognlogshard} but for the 0.5--2.0\keV~band. The open circles are from the ultradeep (1 Msec) ROSAT HRI observation in the Lockman Hole (Hasinger et al.~1998), and the solid points are from shallower ROSAT surveys (Hasinger et al.~1993).}
\label{fig:lognlogssoft}
\end{figure}

\begin{figure}
\psfig{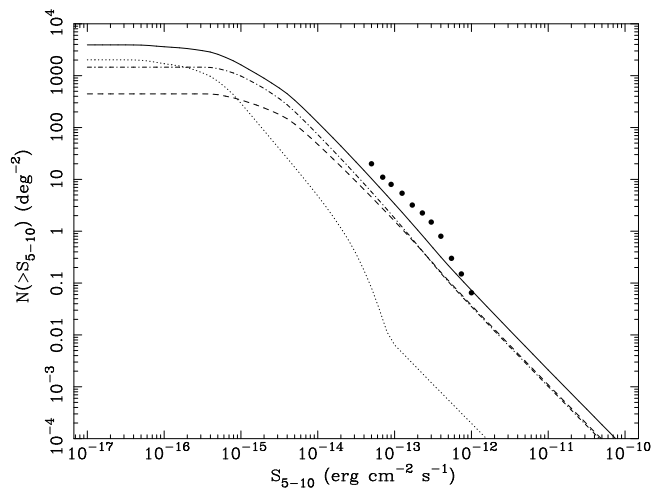}
\caption{\normalsize As in Fig.~\ref{fig:lognlogshard} but for the 5-10\keV~band. The points are preliminary BeppoSAX HELLAS results (Comastri et al.~1999).}
\label{fig:lognlogs5to10}
\end{figure}

\begin{figure}
\psfig{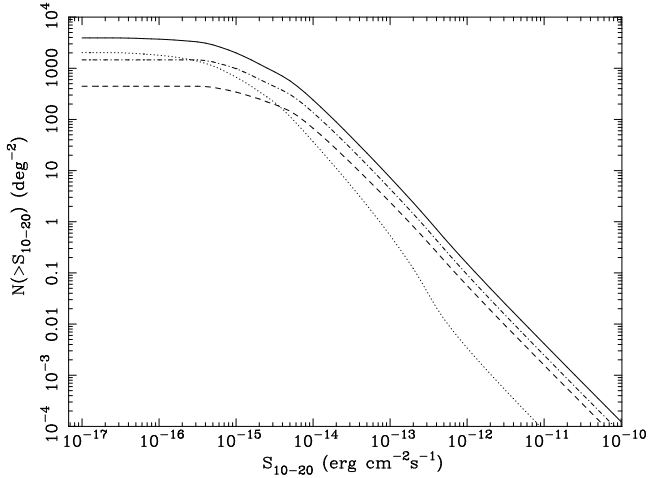}
\caption{\normalsize As in Fig.~\ref{fig:lognlogshard} but for the 10-20\keV~band, potentially important for future missions.}
\label{fig:lognlogs10to20}
\end{figure}

\begin{figure}
\psfig{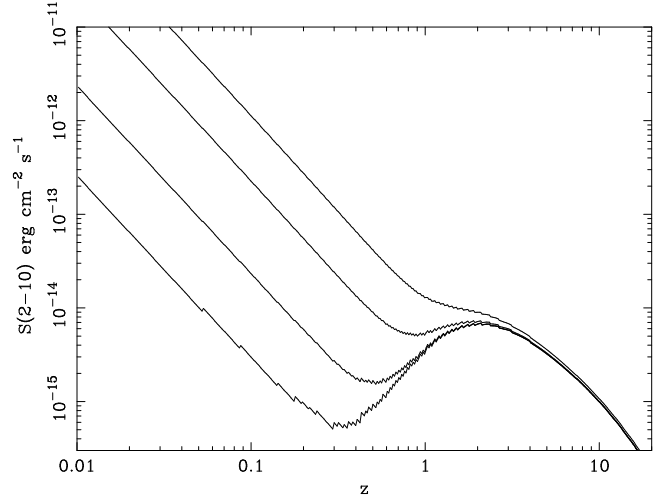}
\caption{\normalsize For a completely unabsorbed (rest-frame) 2--10\keV~luminosity of $10^{45}$\ergps, the curves show the observed 2--10\keV~flux as a function of redshift from a source with $N_{\rm{H}}=10^{24.5}$\pcmsq, for scattering fractions of (from top to bottom) 5, 1, 0.1 and 0.01 per cent, respectively. The strong negative K-correction acting upon the 30\keV~rest-frame peak is clearly evident.}
\label{fig:negkcorr}
\end{figure}

The `kink' which appears at a flux of $10^{-13}$\ergpcmsqps in the contribution of Compton-thick type 2 sources to Fig.~\ref{fig:lognlogshard} is due to a negative K-correction which applies to the spectra of these sources (a similar feature is present in the Compton-thin type 2 contribution to Fig.~\ref{fig:lognlogssoft}). Indeed, Fig.~\ref{fig:mcspectraFe5} shows that the spectrum of a source with $N_{\rm{H}}\sim10^{24.5}$\psqcm~rises very steeply above 10\keV, and the redshifting of this rest-frame 30\keV~peak down into the observed 2--10\keV~band can thus more than offset the $D_{\rm{L}}^{-2}$ cosmological dimming. The extent of the effect is illustrated in Fig.~\ref{fig:negkcorr}, where we show the observed flux from such a source as a function of redshift, for several values of the scattering fraction. It is thus apparent that deep Chandra and XMM pointings will be able to detect the high luminosity end [$L(2-10\keV) \geq 10^{45}\ergps$ in the unabsorbed rest-frame] of the XLF for $N_{\rm{H}}\leq 10^{25}$\psqcm~(above which $N_{\rm{H}}$ the K-correction is much smaller) out to the highest redshifts at which such sources could plausibly exist. This is analogous to the way in which the sub-mm-detectability of star-forming galaxies with Arp 220-like rest-frame far-infrared SEDs is greatly enhanced between redshifts 2--5 (Blain \& Longair 1993).

We note that the models are much less sensitive to the value of $z_{max}$ than they are to either $p$ or $z_{cut}$. The existence of sources at $z \gg z_{max}$ thus cannot be ruled out.

\section{SUMMARY AND CONCLUSIONS} 
We have re-examined AGN synthesis models for the XRB in the light of evidence suggesting that more than half of all Seyfert 2 galaxies in the local universe are Compton-thick (Risaliti et al.~1999). Compton down-scattering and the iron abundance of the absorbing material influence the appearance of the source spectra around 30\keV~in ways which significantly affect the fit to the XRB spectrum, as summarised below.

(1) It is possible to fit the XRB spectrum assuming a solar iron abundance, but there is no scope for accommodating an increased type 2/type 1 ratio at higher redshifts.

(2) With an iron abundance of $5\times$~solar, the fit to the XRB spectrum is better and much more parameter space becomes available. We illustrated this with an arbitrarily chosen model in which the assumed type 2/type 1 ratio increases from a local value of 3.8 to 9.7 above $z=2$. This model predicts that the forthcoming X-ray missions will uncover type 2 sources in substantial numbers below $S(2-10\keV)=10^{-14}$\ergpcmsqps. Furthermore, a strong negative K-correction facilitates the detection of Compton-thick sources with $10^{24} \leq N_{\rm{H}} \leq 10^{25}$\pcmsq~out to high redshift.

As our aim was to demonstrate what is possible within the models, we do not discuss the cosmological constraints bearing upon the existence of large iron abundances in high-redshift AGN. We note though that the strength of the NV$\lambda1240$ emission-line in some quasars demands metallicities of at least 5$Z_{\rm{\odot}}$ (Ferland et al.~1996), consistent with models of early-epoch rapid star formation. In a recent review, Hamann and Ferland~(1999) concluded that there is a growing consensus for super-solar metallicities in quasars out to $z>4$ (typically at the level of one to a few times solar), with the suggestion of higher metallicities in more luminous quasars and with no evidence for a decline at the highest redshifts. Concerning iron itself, they noted that a super-solar Fe/Mg abundance out to $z>4$ is one possible solution to an energy budget problem with the broad FeII emission lines. We also note that studies of the fluorescent iron K$\alpha$ line in some Seyfert 1 galaxies [see eg. Reynolds et al.~(1995) and Lee et al.~(1999)] are consistent with iron abundances of $\sim 2\times$~solar. 

A super-solar metallicity can, however, plausibly be accommodated within the absorption model of Fabian et al.~(1998) for the sources which constitute the XRB. The latter authors noted the need for an obscuring medium with a large covering fraction as seen from the central source, in order to generate the high fraction of obscured AGN which early XRB models recognised as being necessary (Setti \& Woltjer~1989). They thus proposed that nuclear starbursts obscure the XRB, with the continual injection of energy from supernovae serving to keep the medium well `inflated'. Without this mechanism it would rapidly dissipate into the `standard' dusty torus invoked by unified schemes, leading to too low a covering fraction. These same supernovae could supply and subsequently maintain (by, for example, preventing depletion onto dust grains) a high gas phase iron abundance. A high metallicity would also reduce the mass of gas required for a given amount of absorption.

\section*{ACKNOWLEDGMENTS} RJW acknowledges a PPARC Studentship and ACF thanks the Royal Society for support. 

{}

\end{document}